# Portfolio Rebalancing under Uncertainty Using Meta-heuristic Algorithm

Mostafa Zandieh [a,*] , Seyed Omid Mohaddesi [b]

[a] Department of Industrial Management, Management and Accounting Faculty, Shahid Beheshti University, Tehran, Iran

[b] Department of Financial Engineering, Raja University, Qazvin, Iran

**Abstract**

In this paper, we solve portfolio rebalancing problem when security returns are represented by uncertain variables considering transaction costs. The performance of the proposed model is studied using constant-proportion portfolio insurance (CPPI) as rebalancing strategy. Numerical results showed that uncertain parameters and different belief degrees will produce different efficient frontiers, and affect the performance of the proposed model. Moreover, CPPI strategy performs as an insurance mechanism and limits downside risk in bear markets while it allows potential benefit in bull markets. Finally, using a globally optimization solver and genetic algorithm (GA) for solving the model, we concluded that the problem size is an important factor in solving portfolio rebalancing problem with uncertain parameters and to gain better results, it is recommended to use a meta-heuristic algorithm rather than a global solver.

## 1. Introduction

Portfolio rebalancing is one of the major components of portfolio management process [1]. Generally, asset allocation is performed by investors in accordance with their level of risk seeking or risk aversion as well as their expected portfolio return. Meanwhile, the objective of portfolio rebalancing is to restore a portfolio to its original state and primarily optimal allocation. Portfolio rebalancing is considered to be one of the risk control techniques; since as time passes, changes in asset prices lead to gains and losses on each asset in portfolio which will cause an increase or decrease in the weight invested in that asset, and as a result the risk of investment will be increased. This will make the portfolio to deviate beyond a certain threshold from the investor's expected risk [2]. Portfolio rebalancing is a dynamic process of buying and selling of assets in portfolio so that the optimal weights invested in each asset is maintained over time.

In this situation considering transaction cost associated with buying and/or selling of an asset is important. In fact, transaction cost is one of the main constraints for modeling portfolio rebalancing that help in creating more realistic models [3]. Many researchers studied portfolio rebalancing problem considering transaction costs including Sun et al. [4], 2006, Fang et al. [5], 2006, Fadaei-Nezad and Banaeian [6], 2010, Yu and Lee [7], 2011, Woodside-Oriakhi et al. [8], 2013, Gupta et al. [9], 2014, Wang et al. [10], 2014, Chen et al. [11], 2014, Qin et al. [12], 2014, Rabbani [13], 2014 and Kumar et al. [14], 2015. In order to measure portfolio performance different scholars practiced various models considering different parameters including risk, return, liquidity, investment horizon and so forth. To the best of our knowledge, however, all solved their models assuming buy-and-hold (B&H) as rebalancing strategy.

---

* Corresponding Author. Phone: +98 (21) 29905215; Mobile: +98 9123588698;
E-mail address: *m_zandieh@sbu.ac.ir*

In addition, the most notable characteristic of security returns is uncertainty [15]. In classical portfolio theory, security returns were described by random variables, and back then probability theory was the main mathematical tool for handling uncertainty. However, uncertainty is varied in real and complex world, especially when human factors are involved, and randomness is not the only type of uncertainty in reality. Consequently, security market as one of the most complex markets in the world, contains almost all kinds of uncertainty. In particular, the security returns are sensitive to various factors including economic, social, and political, and very importantly, people's psychological factors. Due to this complex nature of financial markets, historical data may be insufficient to reflect the future returns of securities in real situations [12]. Another feasible approach for estimating probability distribution of security returns is using belief degrees evaluated by experts. To deal with this belief degrees, Liu [16] founded the concept of uncertain measure and uncertainty theory. Liu also proposed uncertain programming for solving optimization problems involving uncertain variables. In this area, there have been many studies among which we can refer to vehicle routing and project scheduling problems, shortest path problem and stock model [17]. In particular, Yan and Huang [15, 18] applied Liu's uncertainty theory to portfolio selection problem using uncertain variables when securities returns are neither random nor fuzzy. Moreover, in order to solve uncertain portfolio optimization model, Zhang et al. and Chen proposed meta-heuristic algorithms [19, 20]. Yet there aren't too many studies in the literature on portfolio rebalancing under uncertainty using experts' subjective evaluations. One of the few studies on uncertain portfolio rebalancing model is the one done by Qin et al. [12] in 2014. They used buy-and-hold strategy and solved their proposed model by numerical examples for small size problems. Table 1 represents the gap analysis in the literature of the portfolio rebalancing and the status of the present study in the context. According to the type of mathematical programming model, considered parameters and main constraints of the model, types of variables and the rebalancing strategy, we categorized the aforementioned works which clearly indicates the gap in the context.

**Table 1.** Analysis of gap in the literature

| Ref. | Model | | | Parameters and Constraints | | | | | Type of Variables | | | Rebalancing Strategy | | |
|---|---|---|---|---|---|---|---|---|---|---|---|---|---|---|
| | Multi-objective | Markowitz (Mean-Variance) | Markowitz Multi-period Other | Risk | Return | Liquidity | Investment Horizon | Transaction Costs | Fuzzy | Random | Uncertain | Buy & Hold | CPPI | Other (Innovative) |
| [3] | ✓ | | | ✓ | ✓ | ✓ | | ✓ | ✓ | | | ✓ | | |
| [4] | ✓ | | | ✓ | ✓ | ✓ | | ✓ | ✓ | | | ✓ | | |
| [5] | | | ✓ | ✓ | ✓ | | | ✓ | | ✓ | | | | ✓ |
| [6] | ✓ | | | ✓ | ✓ | ✓ | | | ✓ | | | ✓ | | |
| [7] | ✓ | | | ✓ | ✓ | ✓ | | ✓ | ✓ | | | ✓ | | |
| [8] | | ✓ | | ✓ | ✓ | | ✓ | ✓ | | ✓ | | ✓ | | |
| [9] | ✓ | | | ✓ | ✓ | ✓ | | ✓ | ✓ | | | ✓ | | |
| [10] | ✓ | | | ✓ | ✓ | ✓ | | ✓ | ✓ | | | ✓ | | |
| [11] | | ✓ | | ✓ | ✓ | | | ✓ | ✓ | | | ✓ | | |
| [12] | | ✓ | | ✓ | ✓ | | | ✓ | | | ✓ | ✓ | | |
| [13] | | | ✓ | ✓ | ✓ | | ✓ | ✓ | | ✓ | | ✓ | | |
| [14] | ✓ | | | ✓ | ✓ | | | ✓ | | ✓ | | ✓ | | |
| [21] | | | ✓ | ✓ | ✓ | | | ✓ | | ✓ | | ✓ | | |
| [22] | ✓ | | | ✓ | ✓ | | ✓ | ✓ | | ✓ | | ✓ | | |
| [23] | | | ✓ | ✓ | ✓ | ✓ | | ✓ | ✓ | | | ✓ | | |
| **Current Study** | ✓ | | | ✓ | ✓ | | | ✓ | | | ✓ | ✓ | ✓ | |

Accordingly, the contribution of our study is first to integrate constant-proportion portfolio insurance (CPPI) strategy with uncertain variables in portfolio rebalancing problem. Second, the results of solving our integrated model is compared with a corresponding uncertain model with buy-and-hold (B&H) strategy. Third, the portfolio rebalancing model with CPPI strategy is solved using a meta-heuristic algorithm which has been never investigated in the context.

The objective of this paper is to study portfolio rebalancing problem using uncertain variables when security returns are estimated by experts' evaluated belief degrees. Here, we attempt to solve the model using constant-proportion portfolio insurance (CPPI) as rebalancing strategy for different size problems. The performance of CPPI strategy and B&H strategy is compared by solving the model using real data from Tehran Stock Exchange (TSE). In addition, a genetic algorithm is proposed to solve the uncertain portfolio rebalancing model and its effectiveness is measured and illustrated by solving real examples for different size problems.

The rest of the paper is organized as follows. In Section 22, we review some basic definitions and fundamentals of uncertainty theory and portfolio rebalancing. Section 3 formulates the portfolio rebalancing model when security returns are described by uncertain variables. In Section 4, we present details of the solution algorithm for solving the model. The model is tested in Section 5 using 11 problems in three different sizes considering 100 companies selected from Tehran Stock Exchange. This section also discusses the obtained results. Finally, we conclude the paper and provide some suggestions for future researches in Section 6.

## 2. Preliminaries

Uncertainty theory was first introduced by Liu [16] in 2007 and further developed by other researchers. Today uncertainty theory is considered as a branch of axiomatic mathematics for modeling belief degrees. Here we review some basic concepts, definitions and properties of uncertainty theory including uncertain measure, uncertain variable and uncertainty distribution, which will be used in the whole paper.

An uncertain measure $\mathcal{M}$ on the $\sigma$-algebra $\mathcal{L}$ is defined as a number $\mathcal{M}\{\Lambda\}$ which is assigned to each event $\Lambda$ to indicate the belief degree with which we believe $\Lambda$ will happen. Obviously the assignment of this number is not arbitrary, and the uncertain measure $\mathcal{M}$ must have certain mathematical properties. To insure this and in order to rationally deal with belief degrees, Liu [16] proposed the following three axioms:

- **Axiom 1.** $\mathcal{M}\{\Gamma\} = 1$.
- **Axiom 2.** $\mathcal{M}\{\Lambda\} + \mathcal{M}\{\Lambda^c\} = 1$ for any event $\Lambda$.
- **Axiom 3.** For every countable sequence of events $\Lambda_1, \Lambda_2, \ldots$ we have

$$\mathcal{M}\left\{\bigcup_{i=1}^{\infty} \Lambda_i\right\} \leq \sum_{i=1}^{\infty} \mathcal{M}\{\Lambda_i\}. \quad (1)$$

Product uncertain measure was defined by Liu [24] in 2009 and produced the forth axiom of uncertainty theory.

- **Axiom 4.** Let $(\Gamma_k, \mathcal{L}_k, \mathcal{M}_k)$ be uncertainty spaces for $k = 1,2,\ldots$; The product uncertain measure $\mathcal{M}$ is an uncertain measure satisfying

$$\mathcal{M}\left\{\prod_{k=1}^{\infty} \Lambda_k\right\} = \bigwedge_{k=1}^{\infty} \mathcal{M}_k\{\Lambda_k\} \quad (2)$$

where $\Lambda_k$ are arbitrarily chosen events from $\mathcal{L}_k$ for $k = 1,2,\ldots$, respectively.

Uncertain variable is a fundamental concept in uncertainty theory [17], and is defined as a function $\xi$ from an uncertainty space $(\Gamma, \mathcal{L}, \mathcal{M})$ to the set of real numbers such that $\{\xi \in B\}$ is an event for any Borel set $B$. In order to describe uncertain variable $\xi$ Liu introduced uncertainty distribution $\Phi$ which is defined by

$$\Phi(x) = \mathcal{M}\{\xi \leq x\} \tag{3}$$

for any real number $x$.

For example, a linear uncertain variable is an uncertain variable $\xi$ in which has a linear uncertainty distribution

$$\Phi(x) = \begin{cases} 0, & if\ x \leq a, \\ \dfrac{x-a}{b-a}, & if\ a \leq x \leq b, \\ 1, & if\ x \geq b, \end{cases} \tag{4}$$

denoted by $\mathcal{L}(a,b)$ where $a$ and $b$ are real numbers with $a < b$. A normal uncertain variable is an uncertain variable $\xi$ in which has a normal uncertainty distribution

$$\Phi(x) = \left(1 + \exp\left(\frac{\pi(e-x)}{\sqrt{3}\sigma}\right)\right)^{-1}, \quad x \in \Re \tag{5}$$

denoted by $\mathcal{N}(e,\sigma)$ where $e$ and $\sigma$ are real numbers with $\sigma > 0$.

In addition to uncertainty distribution, a function $\Phi^{-1}$ is called an inverse uncertainty distribution of an uncertain variable $\xi$ if and only if $\mathcal{M}\{\xi \leq \Phi^{-1}(\alpha)\} = \alpha$, for all $\alpha \in [0,1]$. Liu [16] also stated that uncertain variables $\xi_1, \xi_2, \ldots, \xi_n$ are considered independent if for any Borel sets $B_1, B_2, \ldots, B_n$ we have

$$\mathcal{M}\left\{\bigcap_{i=1}^{n}(\xi_i \in B_i)\right\} = \bigwedge_{i=1}^{n} \mathcal{M}\{\xi_i \in B_i\} \tag{6}$$

In order to represent the size of uncertain variable, Liu [16] proposed the expected value of $\xi$ as

$$E[\xi] = \int_0^{+\infty} \mathcal{M}\{\xi \geq x\}\,dx - \int_{-\infty}^{0} \mathcal{M}\{\xi \leq x\}\,dx \tag{7}$$

provided that at least one of the two integrals is finite.

For example, the linear uncertain variable $\xi \sim \mathcal{L}(a,b)$ has an expected value $E[\xi] = (a+b)/2$. The normal uncertain variable $\xi \sim \mathcal{N}(e,\sigma)$ has an expected value $e$, which means $E[\xi] = e$.

**Lemma 1** [24] Let $a$ and $b$ be two real numbers, and $\xi$ and $\eta$ two uncertain variables. Then we have $E[a\xi + b] = aE[\xi] + b$. Further, if $\xi$ and $\eta$ are independent, then $E[a\xi + b\eta] = aE[\xi] + bE[\eta]$.

A degree of the spread of the distribution around its expected value is how Liu defined the variance of uncertain variable [17]. Let $\xi$ be an uncertain variable with finite expected value $e$. Then the variance of $\xi$ is $V[\xi] = E[(\xi - e)^2]$. This means that the variance is the expected value of $(\xi - e)^2$, and owing to the fact that $(\xi - e)^2$ is a nonnegative uncertain variable, we also have

$$V[\xi] = \int_0^{+\infty} \mathcal{M}\{(\xi - e)^2 \geq x\}\,dx \tag{8}$$

For example, the linear uncertain variable $\xi \sim \mathcal{L}(a,b)$ has the variance $V[\xi] = (b-a)^2/12$, and the normal uncertain variable $\xi \sim \mathcal{N}(e,\sigma)$ has the variance $\sigma^2$.

**Lemma 2** [24] Let $a$ and $b$ be real numbers, and $\xi$ an uncertain variable with finite expected value, then $V[a\xi + b] = a^2 V[\xi]$. Further, let $e$ be the expected value of uncertain variable $\xi$. then $V[\xi] = 0$ if and only if $\mathcal{M}\{\xi = e\} = 1$. This means the uncertain variable $\xi$ is basically the constant $e$.

**Lemma 3** [17] Let $\xi$ be an uncertain variable with regular uncertainty distribution $\Phi$ and finite expected value $e$. Then

$$V[\xi] = \int_0^1 (\Phi^{-1}(\alpha) - e)^2\,d\alpha \tag{9}$$

**Lemma 4** [17] Assume $\xi_1, \xi_2, \ldots, \xi_n$ are independent uncertain variables with regular uncertainty distributions $\Phi_1, \Phi_2, \ldots, \Phi_n$, respectively. If $f(x_1, x_2, \ldots, x_n)$ is strictly increasing for $x_1, x_2, \ldots, x_m$ and strictly decreasing for $x_{m+1}, x_{m+2}, \ldots, x_n$, then expected value and variance of the uncertain variable $\xi = f(\xi_1, \xi_2, \ldots, \xi_n)$ are as follows

$$E[\xi] = \int_0^1 f\left(\Phi_1^{-1}(\alpha), \ldots, \Phi_m^{-1}(\alpha), \Phi_{m+1}^{-1}(1-\alpha), \ldots, \Phi_n^{-1}(1-\alpha)\right) d\alpha \tag{10}$$

$$V[\xi] = \int_0^1 \left(f\left(\Phi_1^{-1}(\alpha), \ldots, \Phi_m^{-1}(\alpha), \Phi_{m+1}^{-1}(1-\alpha), \ldots, \Phi_n^{-1}(1-\alpha)\right) - e\right)^2 d\alpha \tag{11}$$

## 2.1. Rebalancing strategy

Individual and institutional investors use various strategies for rebalancing their portfolio in order to optimize their investment process. Perold and Sharpe [25] introduced four dynamic strategies for portfolio rebalancing which include: 1) buy-and-hold; 2) constant mix; 3) constant-proportion portfolio insurance (CPPI); and 4) option-based portfolio insurance (OBPI). As mentioned earlier, no scholar has solved the portfolio rebalancing problem using uncertain variables and CPPI strategy to date. In this paper, we use CPPI strategy to solve portfolio rebalancing problem while uncertain variables of the model are estimated by experts' evaluated belief degrees.

### 2.1.1. Constant-proportion portfolio insurance

The CPPI strategy is a self-financing dynamic strategy in which by investing a portion of the wealth in risky asset equal to a constant multiple of cushion, the investor limits the downside risk while gaining some upside potential [26]. The cushion is equal to the difference between value of the portfolio at time $t$ ($W_t$) and the floor which the investor refuses the value of the portfolio to go below ($F_t$), and is defined as

$$C_t = W_t - F_t \tag{12}$$

In this case and at any time $t$, if $W_t > F_t$, the exposure to the risky asset (the amount of wealth invested in risky asset) is obtained by $mC_t \equiv m(W_t - F_t)$, where $m > 1$ is a constant multiplier. If $W_t \leq F_t$, the entire portfolio is invested in risk-free bonds. In fact, CPPI always keeps a constant multiple of the cushion as exposure to risky asset like stocks, and the rest of the wealth is invested in risk-free government savings or treasury bonds [27]. Thus, if $E_t$ represents the exposure to the risky asset, and $E_t/W_t$ a fraction of the total wealth to be invested in risky assets at any time $t$, then

$$\frac{E_t}{W_t} = \min\left\{m\left(1 - \frac{F_t}{W_t}\right), 1\right\} \tag{13}$$

where $m > 1$ is a constant multiplier. The value of the multiplier $m$ is derived based on the investor's risk seeking or risk aversion, and typically by answering what is the probability of the maximum one-day loss from investing in risky asset. The multiplier will be the inverse of this percentage of the loss. For instance, if an investor estimates the maximum probable loss of 20%, the multiplier value will be equal to 5. Values between 3 and 6 are more frequently used as multiplier [26].

## 3. Uncertain portfolio rebalancing model

In this section, the bi-objective rebalancing problem is formulated using uncertain variables and considering transaction costs. As previously mentioned, in this paper we use CPPI strategy for modeling portfolio rebalancing problem. For this purpose, we assume that the investors allocate their wealth in accordance with CPPI strategy assumptions, and by dividing it between $n$ risky assets (stocks) and 1 risk-free asset (participation bonds in this paper). The parameters and variables used to formulate the mathematical model are described as follows:

$\xi_i$: the uncertain return of the $i$-th asset,
$u_i$: the maximal fraction of the wealth allocated to the $i$-th asset,
$l_i$: the minimal fraction of the wealth allocated to the $i$-th asset,

$x_i^0$: the initial proportion of the total funds invested in the $i$-th asset, i.e., before rebalancing,
$x_i^+$: the proportion of the $i$-th asset to be purchased during rebalancing,
$x_i^-$: the proportion of the $i$-th asset to be sold during rebalancing,
$x_i$: the final proportion of the total funds invested in the $i$-th asset, i.e., after rebalancing,
$y_i$: a binary variable indicating whether the $i$-th asset is contained in the portfolio, where
$$y_i = \begin{cases} 1, & \text{if the } i-\text{th asset is contained in the portfolio} \\ 0, & otherwise, \end{cases}$$
$z_i$: a binary variable indicating whether the $i$-th asset is purchased or sold, where
$$z_i = \begin{cases} 1, & \text{if the } i-\text{th asset is purchased,} \\ 0, & \text{if the } i-\text{th asset is sold,} \end{cases}$$
$h$: the maximum number of assets in the portfolio,
$b_i$: the transaction cost of buying a proportion of the $i$-th asset,
$s_i$: the transaction cost of selling a proportion of the $i$-th asset,
$C(x)$: the total transaction costs incurred by rebalancing the portfolio,
$W_t$: the total value of the portfolio (entire wealth) at time $t$,
$F_t$: the minimum value accepted by the investor, which they refuse the value of the portfolio to go below,
$m$: the constant multiplier of CPPI strategy.

**Agreement** The $zeroth$ asset (i.e., $x_0$) will indicate the risk-free asset (i.e., participation bonds) in rest of the paper, thus we also have the following notations

$x_0^0$: the initial proportion of the total funds invested in the risk-free asset, i.e., before rebalancing,
$x_0^+$: the proportion of the risk-free asset to be purchased during rebalancing,
$x_0^-$: the proportion of the risk-free asset to be sold during rebalancing,
$x_0$: the final proportion of the total funds invested in the risk-free asset, i.e., after rebalancing.

### 3.1. Transaction costs

Transactions cost is one the principal constraints of portfolio rebalancing problem which helps in creating more realistic models by integrating market frictions [3]. In fact, transaction costs make a decrease in portfolio return; hence, considering these costs is sensible for portfolio rebalancing as well as portfolio optimization problems.

We assume that $x^0$ stands for the initial portfolio and our goal is to achieve optimal portfolio $x$ by adjusting the weights of each asset in portfolio. In this case, transaction costs on purchases are measured by the amount added to the portfolio $x^0$, and transaction costs on sales are measured by the amount deducted from the initial portfolio $x^0$. If the investor pays transaction costs proportional to $b_i$ for every added amount to the $i$-th asset ($x_i^+$), and proportional to $s_i$ for every deducted amount from the $i$-th asset ($x_i^-$), the total transaction costs of rebalancing the portfolio will be derived by

$$C(x) = \sum_{i=0}^{n}(b_i x_i^+ + s_i x_i^-) \qquad (14)$$

where for $i = 1,2, \ldots, n$, $s_i$ and $b_i$ indicate transaction costs of stocks (risky assets), while $s_0$ and $b_0$ indicate transaction costs of participation bonds (risk-free asset); and are obtained by Table 2 and Table 3, respectively.

### 3.2. Objectives

The proposed model is a bi-objective optimization model in which the returns on the assets are considered as uncertain variables. The objectives of the model will be described below.

- **Objective 1: Portfolio return**

The main goal of investment is to produce return. According to Markowitz mean-variance portfolio theory [28], the investment return can be modeled by means of expected return. Thus, the first objective of our proposed model, is to maximize portfolio return calculated by uncertain expected value. In addition, the transaction costs need to be considered in objective function as a factor making a decrease in portfolio return.

Assume the returns of $n$ assets are represented by uncertain variables $\xi_i$ for $i = 0,1,2,\ldots,n$. The objective function to maximize portfolio return after adjusting transaction costs is expressed as

$$\boldsymbol{max}\, E\left[\sum_{i=0}^{n} x_i \xi_i\right] - C(x) \qquad (15)$$

where $x_i$ indicates the proportion of total funds invested in the $i$-th asset and $C(x)$ is calculated by Equation (14).

Table 2. Transaction commission & fees for stocks in Tehran Stock Exchange

| Fee Description | The Buyer Fees (%) | The Seller Fees (%) |
|---|---|---|
| Brokerage Fees | 0.4% | 0.4% |
| TSE Commission | 0.032% | 0.048% |
| SEO Commission | 0.032% | 0.048% |
| Clearing Fees | 0.022% | 0.033% |
| Taxes | - | 0.5% |
| Total Commission & Fees | 0.486% | 1.029% |

Table 3. Transaction commission & fees for participation bonds

| Fee Description | The Buyer Fees (%) | The Seller Fees (%) |
|---|---|---|
| Brokerage Fees | 0.063% | 0.063% |
| TSE Commission | 0.0096% | 0.0144% |
| Total Commission & Fees | 0.0726% | 0.0774% |

- **Objective 2: Portfolio risk**

Huang [15] describes the optimal portfolio as the one minimizing the risk while maximizing the return. To be more precise, when higher levels of returns won't be attained unless we take more risk, or taking less risk won't be possible unless we undertake lower levels of return, we will achieve the optimal portfolio. According to Markowitz model [28], the risk of the portfolio is measured by means of variance. Let $\xi_i$, $i = 0,1,2,\ldots,n$ be the uncertain returns of $n$ assets. Then the objective function to minimize the risk of the portfolio is described as

$$\boldsymbol{min}\, V\left[\sum_{i=0}^{n} x_i \xi_i\right] \qquad (16)$$

where $x_i$ indicates the proportion of total funds invested in the $i$-th asset.

### 3.3. Constraints

- *Rebalancing constraint*: The final proportion of the total funds invested in the $i$-th asset after rebalancing is calculated by

$$x_i = x_i^0 + x_i^+ - x_i^-, \qquad i = 0,1,2,\ldots,n. \qquad (17)$$

- *Complementarity constraint on buying and selling an asset*: In any timeframe, the investor must either buy a proportion of the $i$-th asset or sell it. In other words, it is impossible to buy and sell some proportion of the same asset simultaneously [3]. Thus, $x_i^+$ and $x_i^-$ are complementary and the following constraint must hold

$$x_i^+ \cdot x_i^- = 0, \qquad i = 0,1,2,\ldots,n. \qquad (18)$$

In order to better control the above constraint in our model, it is transformed to the linear form by replacing with the following two constraints

$$x_i^+ \leq z_i, \qquad i = 0,1,2,\ldots,n, \tag{19}$$

$$x_i^- \leq (1 - z_i), \qquad i = 0,1,2,\ldots,n, \tag{20}$$

where $z_i \in \{0,1\}$.

- *Capital budget constraint*: In the proposed model, we assume that the portfolio rebalancing process is financed by current assets in the portfolio (i.e., the portfolio is self-financing), and the investor does not add any additional capital to the portfolio during rebalancing. The transaction costs are also paid by the capital in the portfolio. Moreover, since we use CPPI strategy for rebalancing, the exposure is derived by Equation (13). Thus, the following constraint must hold

$$\sum_{i=1}^{n} x_i + C(x) = \min\left\{m\left(1 - \frac{F_t}{W_t}\right), 1\right\}. \tag{21}$$

- *Risk-free asset constraint*: The amount of investment in risk-free asset (participation bonds) is equal to the subtraction of total wealth and the exposure to the risky asset. Thus this amount is calculated by

$$x_0 = 1 - \min\left\{m\left(1 - \frac{F_t}{W_t}\right), 1\right\}. \tag{22}$$

- *Maximal and minimal fraction of the wealth allocated to one asset*: The maximal and minimal fraction of the wealth that can be allocated to a specific asset in the portfolio may depend on several factors [3]. For example, the investor may consider the price or value of the asset to the average price or value of all assets in the portfolio, the minimum volume that can be ordered and traded in the market, the past behavior of the price or traded volume of the asset, the available information about the issuer of the asset, or the trends in a particular industry. Generally, the constraints corresponding to lower bounds ($l_i$) and upper bounds ($u_i$) on investment in a particular asset, are added to the portfolio rebalancing model in order to avoid a large number of low volume investments and also to ensure adequate diversification in portfolio. Thus, the following constraints must hold

$$x_i \leq u_i y_i, \qquad \forall i = 0,1,2,\ldots,n \tag{23}$$

$$x_i \geq l_i y_i, \qquad \forall i = 0,1,2,\ldots,n \tag{24}$$

where $(0 \leq l_i \leq u_i \leq 1, \forall i)$.

- *Number of assets*: In order to effectively manage risky assets in the portfolio, and to avoid large number of low volume investments, the following constraint is defined for managing the number of risky assets.

$$\sum_{i=1}^{n} y_i \leq h. \tag{25}$$

where $y_i \in \{0,1\}$.

- *No short selling constraint*: In order to avoid short selling of assets the following three constraints must hold

$$x_i \geq 0, \ x_i^+ \geq 0, \ x_i^- \geq 0, \quad i = 0,1,2,\ldots,n. \tag{26}$$

### 3.4. The decision problem

The bi-objective uncertain portfolio rebalancing model is formulated as follows

$$\begin{cases} \max f_1(x) = E\left[\sum_{i=0}^{n} x_i\xi_i\right] - C(x) \\ \min f_2(x) = V\left[\sum_{i=0}^{n} x_i\xi_i\right] \\ \text{Subject to:} \\ \quad x_i = x_i^0 + x_i^+ - x_i^-, \quad i = 0,1,2,\dots,n, \\ \quad x_i^+ \leq z_i, \quad i = 0,1,2,\dots,n, \\ \quad x_i^- \leq (1 - z_i), \quad i = 0,1,2,\dots,n, \\ \quad \sum_{i=1}^{n} x_i + C(x) = \min\left\{m\left(1 - \frac{F_t}{W_t}\right), 1\right\}, \\ \quad x_0 = 1 - \min\left\{m\left(1 - \frac{F_t}{W_t}\right), 1\right\}, \\ \quad \sum_{i=1}^{n} y_i \leq h, \\ \quad x_i \leq u_i y_i, \quad i = 0,1,2,\dots,n, \\ \quad x_i \geq l_i y_i, \quad i = 0,1,2,\dots,n, \\ \quad x_i \geq 0, \quad i = 0,1,2,\dots,n, \\ \quad x_i^+ \geq 0 \quad i = 0,1,2,\dots,n, \\ \quad x_i^- \geq 0, \quad i = 0,1,2,\dots,n, \\ \quad y_i, z_i \in \{0,1\}, \quad i = 0,1,2,\dots,n. \end{cases} \quad (27)$$

### 3.5. Solution methodologies

Huang [29] introduced so-called "9999 Method" in order to calculate expected value and variance of the portfolio in mean-variance model when securities returns are described by uncertain variables with different uncertainty distributions.

#### 3.5.1. 9999 Method

Assume $\xi_i$ is an uncertain variable with uncertainty distribution $\Phi_i$, and $k_i$ a positive number for $i = 1,2,\dots,n$, respectively. Let $\Psi_i$ represent the uncertainty distributions of $k_i\xi_i$, $i = 1,2,\dots,n$, respectively. Then, we have

$$\Psi_i^{-1}(\alpha) = k_i\Phi_i^{-1}(\alpha). \quad (28)$$

Now considering $\Psi$ to be the uncertainty distribution of $k_1\xi_1 + k_2\xi_2 + \cdots + k_n\xi_n$, we have

$$\Psi^{-1}(\alpha) = \sum_{i=1}^{n} \Psi_i^{-1}(\alpha) = \sum_{i=1}^{n} k_i\Phi_i^{-1}(\alpha). \quad (29)$$

This means the uncertainty distribution $\Psi$ of $k_1\xi_1 + k_2\xi_2 + \cdots + k_n\xi_n$ can be represented on a computer as shown in

Table **4**.

Suppose $\xi = k_1\xi_1 + k_2\xi_2 + \cdots + k_n\xi_n$ is an uncertain variable. According to Lemma 4 and 9999 Method, the expected value and variance of uncertain variable $\xi$ are as follows

$$E[\xi] = \frac{\sum_{j=1}^{9999} \sum_{i=1}^{n} k_i t_{i/j}}{9999}, \quad (30)$$

$$V[\xi] = \frac{\sum_{j=1}^{9999} \left(\left(\sum_{i=1}^{n} k_i t_{i/j}\right) - E[\xi]\right)^2}{9999}. \quad (31)$$

**Table 4.** Presentation of 9999 Method on computer

| $\alpha_i$ | 0.0001 | 0.0002 | 0.0003 | ... | 0.9999 |
|---|---|---|---|---|---|
| $\Phi_1^{-1}(\alpha_i)$ | $t_{1/1}$ | $t_{1/2}$ | $t_{1/3}$ | ... | $t_{1/9999}$ |
| $\Phi_2^{-1}(\alpha_i)$ | $t_{2/1}$ | $t_{2/2}$ | $t_{2/3}$ | ... | $t_{2/9999}$ |
| $\Phi_3^{-1}(\alpha_i)$ | $t_{3/1}$ | $t_{3/2}$ | $t_{3/3}$ | ... | $t_{3/9999}$ |
| $\vdots$ | $\vdots$ | $\vdots$ | $\vdots$ | $\ddots$ | $\vdots$ |
| $\Phi_n^{-1}(\alpha_i)$ | $t_{n/1}$ | $t_{n/2}$ | $t_{n/3}$ | ... | $t_{n/9999}$ |
| $\Psi^{-1}(\alpha_i)$ | $\sum_{i=1}^{n} k_i t_{i/1}$ | $\sum_{i=1}^{n} k_i t_{i/2}$ | $\sum_{i=1}^{n} k_i t_{i/3}$ | ... | $\sum_{i=1}^{n} k_i t_{i/9999}$ |

As a result, the objectives of the proposed portfolio rebalancing model can be replaced by followings

$$\boldsymbol{max}\, f_1(x) = \frac{\sum_{j=1}^{9999} \sum_{i=0}^{n} x_i t_{i/j}}{9999} - C(x), \tag{32}$$

$$\boldsymbol{min}\, f_2(x) = \frac{\sum_{j=1}^{9999} \left( \left( \sum_{i=0}^{n} x_i t_{i/j} \right) - e \right)^2}{9999}, \tag{33}$$

where $e = \sum_{j=1}^{9999} \sum_{i=0}^{n} x_i t_{i/j} / 9999$ is the expected value of portfolio, and $t_{i/j}$, $i = 0,1,2,...,n$, and $j = 1,2,...,9999$, are obtained from 9999 Method by inverse uncertainty distribution $\Phi_i^{-1}$ according to

Table **4**.

### 3.5.2. ε-constraint method

There are various approaches for solving a multi-objective mathematical programming (MOMP) problem. Miettinen [30] classified them into four categories: 1) no-preferences methods; 2) a priori methods; 3) a posteriori methods; and 4) interactive methods. While in no-preferences methods the decision maker (DM) has no participation in the solution process, a priori methods ask for the DM preferences and opinions before the solution process. In a posteriori methods which are also called generation methods, first the Pareto optimal set (or a representation of it) is generated and then the DM selects the most preferred solution. In interactive methods, the DM gets involved in the solution process by correcting his/her preferences in each iteration and after of being presented only part of the Pareto optimal points.

Considering portfolio selection problems we generally search for every possible combination of assets that generates different efficient portfolios with different combinations of risk – expected return according to the investors' preferences. These different efficient portfolios will form the efficient frontier [29]. Afterwards, each investor can find his/her own optimal portfolio from the efficient frontier according to their risk preferences. Thus, solving the bi-objective portfolio rebalancing problem falls into the a posteriori methods category in which the Pareto optimal set is represented by the efficient frontier, and then the DM (here the investor) will choose the optimal solution according to his/her preferences. Miettinen [30] introduced two basic a posteriori methods; the weighting method and the ε-constraint method. While the ε-constraint method can find every Pareto optimal solution of any MOMP regardless of convexity of the problem, the weighting method fails to find all of the Pareto optimal solutions when the problem is non-convex. Accordingly, we utilize the ε-constraint method in order to solve the bi-objective uncertain portfolio rebalancing problem. In this method, one of the objective functions is optimized by formulating other objective functions as constraints, and transferring them to the constraint part of the model [31]. Assume the following MOMP problem

$$\max \left(f_1(x), f_2(x), \ldots, f_p(x)\right) \tag{34}$$
$$s.t.$$
$$x \in S.$$

Then using ε-constraint method we will have the following single-objective problem

$$\max f_1(x)$$
$$s.t.$$
$$f_2(x) \geq \epsilon_2,$$
$$f_3(x) \geq \epsilon_3, \tag{35}$$
$$\vdots$$
$$f_p(x) \geq \epsilon_p,$$
$$x \in S.$$

The Pareto optimal solutions of the problem are obtained by initialization and parametric variation in the RHS of the constrained objective functions (i.e., $\epsilon_i$) and then solving the model for these different parameters. Consequently, using ε-constraint method, the objective function corresponding to portfolio return in our proposed model is formulated as

$$\frac{\sum_{j=1}^{9999} \sum_{i=1}^{n} x_i t_{i/j}}{9999} - C(x) \geq \lambda, \tag{36}$$

where $\lambda$ represents the minimum expected return required by the investor. By solving the model for different values of $\lambda$, the solutions obtained will form an efficient frontier. In addition, the minimum expected return required by the investor is obviously greater than the return on risk-free asset. Thus

$$\lambda \geq r_f. \tag{37}$$

Finally, the uncertain portfolio rebalancing model is formulated as follows

$$\begin{cases} \min f_2(x) = \dfrac{\sum_{j=1}^{9999}\left(\left(\sum_{i=0}^{n} x_i t_{i/j}\right) - e\right)^2}{9999}, \\ \text{Subject to:} \\ \quad \dfrac{\sum_{j=1}^{9999} \sum_{i=1}^{n} x_i t_{i/j}}{9999} - C(x) \geq \lambda, \\ \quad x_i = x_i^0 + x_i^+ - x_i^-, \quad i = 0,1,2,\ldots,n, \\ \quad x_i^+ \leq z_i, \quad i = 0,1,2,\ldots,n, \\ \quad x_i^- \leq (1 - z_i), \quad i = 0,1,2,\ldots,n, \\ \quad \sum_{i=1}^{n} x_i + C(x) = \min\left\{m\left(1 - \dfrac{F_t}{W_t}\right), 1\right\}, \\ \quad x_0 = 1 - \min\left\{m\left(1 - \dfrac{F_t}{W_t}\right), 1\right\}, \\ \quad \sum_{i=1}^{n} y_i \leq h, \\ \quad \lambda \geq r_f, \\ \quad x_i \leq u_i y_i, \quad i = 0,1,2,\ldots,n, \\ \quad x_i \geq l_i y_i, \quad i = 0,1,2,\ldots,n, \\ \quad x_i \geq 0, \quad i = 0,1,2,\ldots,n, \\ \quad x_i^+ \geq 0 \quad i = 0,1,2,\ldots,n, \\ \quad x_i^- \geq 0, \quad i = 0,1,2,\ldots,n, \\ \quad y_i, z_i \in \{0,1\}, \quad i = 0,1,2,\ldots,n. \end{cases} \tag{38}$$

## 4. Solution Algorithm

Due to the complexity of the nonlinear uncertain portfolio rebalancing problem, it is hard to solve the model using exact methods, and thus, a meta-heuristic algorithm is utilized in this section to solve the proposed model. There are various classifications on the types of meta-heuristics including the type of the search strategy (i.e. local or global), single solution against population-based searches, nature-inspired against non-nature inspired, etc. [32]. For instance, Shahvari and Logendran [33, 34], 2016, and Shahvari et al. [35], 2012, utilized local search-based tabu search (TS) algorithms and tabu search/path relinking (TS/PR) algorithm for their researches. Other researchers employed global search methods that are usually population-based meta-heuristics, such as non-dominated sorting genetic algorithm-II (NSGA-II) and non-dominated ranking genetic algorithms (NRGA) in the work of Sadeghi and Niaki [36], 2015, or particle swarm optimization (PSO) and genetic algorithm (GA) in the works of Mousavi et al. [37, 38], 2014, and Pasandideh et al. [39], 2013, or hybrid algorithm based on GA in the work of Mousavi and Niaki [40], 2013. Furthermore, Mousavi et al. [41-43], 2014 to 2016, used nature-inspired meta-heuristics including fruit fly optimization algorithms (FFOA) and harmony search algorithm.

Selecting the appropriate meta-heuristic algorithm highly depends on the problem itself, and genetic algorithm (GA) has been commonly investigated in the context of portfolio optimization and uncertain programming [3, 19, 44, 45]. As a result, we utilize GA in our study to solve the uncertain portfolio rebalancing problem. Besides, we will use a globally optimization solver based on branch-and-bound concept in order to validate the results obtained and to verify the performance of the proposed GA.

Genetic Algorithm (GA) first became popular through the work of John Holland [46] in early 1970s and has been further developed by others then. The algorithm provides an efficient search method for large spaces that eventually leads to finding the optimal solution. In general, a GA consists of following elements [47]: an encoding mechanism for representing each solution in form of a chromosome, a population of chromosomes, a fitness function assigning scores to each chromosome, genetic operators including selection according to fitness, crossover to produce new offspring and random mutation of new offspring to produce new population. Finally, the evolution process is stopped according to a predetermined termination condition.

### 4.1. Chromosome encoding

In this paper, we consider each chromosome as an array with $n + 1$ elements. Assume $x = (x_0, x_1, x_2, \ldots, x_n)$ is a possible solution of the problem, then for any number $k = 1, 2, \ldots, n + 1$ corresponding to a gene, the gene value represents the final proportion of the wealth invested in $(k - 1)$-th asset.

**Agreement** The first gene of each chromosome represents the final proportion of the wealth invested in risk-free asset (i.e., $x_0$).

#### 4.1.1. Initialization

Initialization of each gene except the first one is taken place by producing uniformly nonnegative numbers between zero and the exposure ($E_t/W_t$). The value of the first gene is obtained by Equation (22). The initialization ensures that constraints corresponding to Equation (26) will hold. In addition, chromosome $x^0$ represents the initial portfolio before rebalancing, and according to Equation (17) we have

$$x_i - x_i^0 = x_i^+ - x_i^-.$$

According to the above equation, subtracting the genes of chromosome $x^0$ from corresponding genes of each chromosome $x_i$ gives us the tradable amount. Thus, Equation (17) will be satisfied. After subtraction, a positive value remaining in each gene indicates adding some proportion to the $i$-th asset, while a negative value indicates selling some proportion from the $i$-th asset. Obviously, the result of the subtraction of chromosomes' genes is either positive or negative, thus the Equation (18) will also be satisfied.

#### 4.1.2. Repairing mechanism

In order to satisfy other constraints and to prevent GA operators from producing infeasible solutions, we need to design some repairing mechanisms. For satisfying constraint of Equation (21), the value of each gene – only genes corresponding to risky assets – need to be adjusted. Thus

$$x'_i = \frac{x_i \times E_t/W_t}{\sum_{i=1}^n x_i} \tag{39}$$

where $x'_i$ is the repaired (adjusted) value of each gene. Moreover, if the maximum number of risky assets ($h$) does not meet in a chromosome after initialization, we will use the following mechanism: 1. Generate a random integer $h'$ between 1 and $h$; 2. Sort genes number 2 to $n+1$ in ascending order based on their values; 3. Replace the value of first $n-h'$ genes with $zero$. Thus, in each chromosome, the maximum number of $h$ genes (assets) have values and the constraint in Equation (25) will be satisfied.

We consider the minimal fraction of the wealth allocated to the $i$-th asset equal to $zero$ in this paper ($l_i = 0$). In order to satisfy the constraint corresponding to maximal fraction of wealth in each asset, the adjusted value of each gene obtained by Equation (39) is compared with predetermined maximal fraction for its corresponding asset ($u_i$), and it will be reduced to $u_i$ if it is greater than its predetermined value. Then, the excess amount of the gene will be added to other nonzero genes. If all genes have their maximum predetermined value, they all will be reduced to $u_i$.

Since $\lambda$ is one of the inputs of the problem, before running the algorithm the constraint of Equation (37) will be controlled already.

### 4.2. Fitness function

The fitness of each chromosome is evaluated by a fitness function. Since all constraints have been satisfied so far leaving the constraint of Equation (36), we consider it while formulating the fitness function. First, we define a penalty function to ensure that Equation (36) is satisfied and to make the actual return of the portfolio exceed minimum expected return. Thus

$$p(x) = \begin{cases} \lambda - (e - C(x)), & if\ \lambda > (e - C(x)) \\ 0, & otherwise, \end{cases} \tag{40}$$

where $e = \sum_{j=1}^{9999} \sum_{i=0}^{n} x_i t_{i/j}/9999$. Incorporating the objective function (33) and the penalty function (40), the fitness function of the algorithm can be defined as follows

$$fitness = \exp\left(-k\big(f_2(x) + M.p(x)\big)\right), \tag{41}$$

where $k$ is a positive constant and $M$ a large positive number. The negative exponent transforms the minimization problem into its equivalent maximization problem for GA to solve. In addition, the exponential function with constant $k$ confines the fitness range and thus alleviates the selection pressure of chromosomes with higher fitness, to prevent the GA from premature convergence [48]. On the other hand, The large positive number $M$ forces the solution to meet constraint (36) before minimizing the portfolio risk.

### 4.3. Genetic operators

- **Crossover operator**

We use roulette wheel method in this paper for selecting chromosomes out of their population in order to perform crossover. For this purpose, we have to select a number of chromosomes from population relative to crossover rate ($p_c$). The probability of the $i$-th chromosome of being selected is equal to

$$p_i = \frac{f_i}{\sum_{i=1}^{N} f_i} \tag{42}$$

where $f_i$ is the fitness value of chromosome $i$ and $N$ is the number of individual chromosomes in population. Parents are randomly and pairwise selected from chosen chromosomes and then the crossover operator is performed to produce new offspring. Here we use one-point crossover. As mentioned earlier, the value of the first gene is equal to the proportion of wealth invested in risk-free asset which is obtained from Equation (22), and therefore it shouldn't be considered during crossover process. Thus, a random number is generated between 2 and the length of the chromosome, then all genes of parent chromosomes are swapped beyond that point.

- **Mutation operator**

Chromosomes selection for mutation operator is totally random and their quantity is determined based on mutation rate ($p_m$). We use two approaches in this paper for random mutation; random swap of genes and random replacement of them. The former includes random selection of two nonnegative genes and swapping them with each other. The latter includes random selection of one gene and changing its value by initialization mechanism mentioned in section 4.1.1. It must be noted that the mutation does not apply to the first gene (i.e., risk-free asset).

- **Elitism**

To maintain and use best solutions in previous generations, we use an elitism operator in which it transfers the best solutions of each iteration to the next generation without any change. The elitism rate is calculated by the following formula

$$p_r = 1 - p_c - p_m. \tag{43}$$

## 5. Computational results

In this section we present numerical examples to test proposed uncertain portfolio rebalancing model and to illustrate corresponding computational results. To do so, we consider three contexts in which we study the performance of the model. First, we study the proposed GA in compare with a globally optimization solver. Then, the performance of CPPI strategy is studied under the proposed uncertain portfolio rebalancing model. Finally, the impact of belief degrees and considering uncertain parameters in the model is investigated.

- **GA performance**

The mathematical model is solved using BARON Solver in GAMS [49]. Moreover, the proposed genetic algorithm is coded and implemented in MATLAB, and on a personal computer with a 2.2 GHz Intel Core i7 CPU and 4 GB of RAM. To generate rebalancing problems, 100 securities are selected from Tehran Stock Exchange (TSE) and using experts' evaluations, normal uncertainty distributions are estimated for the returns on each asset. Eventually, 11 sample problems are divided into 3 different sizes of small, medium and large in order to evaluate the performance of the proposed model. Table 5 represents different sizes of sample problems and corresponding number of assets in each one.

**Table 5.** Problem size and number of assets

| Problem Size | Quantity of Risk-Free Asset(s) | Quantity of Risky Asset(s) |
|---|---|---|
| Small | 1 | 10 |
|  | 1 | 15 |
|  | 1 | 20 |
| Medium | 1 | 30 |
|  | 1 | 40 |
|  | 1 | 50 |
| Large | 1 | 60 |
|  | 1 | 70 |
|  | 1 | 80 |
|  | 1 | 90 |
|  | 1 | 100 |

To determine the feasible region of each problem we first code the uncertain portfolio rebalancing model in GAMS, and by running BARON computational system, a feasible set is specified for $\lambda$. Then by running GA in the search space and for different values of $\lambda$, the corresponding results of 5 times implementation of the algorithm are recorded for each problem. Besides, Figure 1 illustrates an example of GA convergence diagram for the proposed uncertain portfolio rebalancing model.

To compare the results, we calculate the ratio of performance deviation (RPD) for BARON and the proposed GA. RPD represents the superiority of GA results over BARON in which it means the more negative the RPD is, the lower is the risk obtained by GA in compare with BARON. RPD is calculated by the following formula

$$RPD = \frac{Risk_{GA} - Risk_{BARON}}{Risk_{BARON}} \times 100 \tag{44}$$

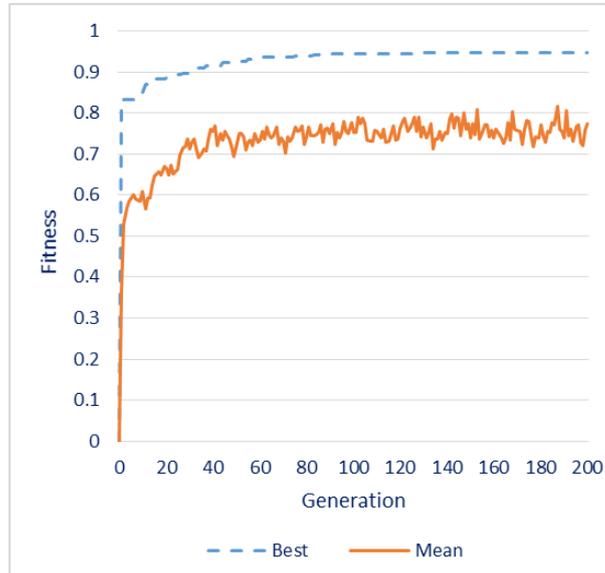

**Figure 1.** Convergence of the GA to maximize the fitness function

Solving the portfolio rebalancing problem with uncertain parameters, we observed that BARON does not guarantee globally optimal solution, not even after finding 20 feasible solutions for each sample problem, and the solver is terminated by reaching its time limit. Table 6 represents the results of solving the model by GA and BARON including average processing time (CPU time) and the corresponding RPD calculated for each sample problem. In addition, for better understanding of results, corresponding efficient frontiers for each sample problem are illustrated in Figure 2, Figure 3 and Figure 4, for small, medium and large size problems, respectively. It can be seen that increasing the problem size, makes the gap between BARON and GA results greater while GA provides better solutions. Moreover, by increasing the problem size, BARON processing time sharply increases while GA reaches better solutions in more reasonable times.

**Table 6.** Results corresponding to GAMS & GA in different problem sizes

| Quantity of Risky Asset(s) | Average Processing Time (Sec) | | RPD% |
|---|---|---|---|
| | BARON | GA | |
| 10 | 368 | 28 | -4% |
| 15 | 386 | 36 | 2% |
| 20 | 287 | 39 | -3% |
| 30 | 540 | 35 | -3% |
| 40 | 774 | 49 | -3% |
| 50 | 823 | 52 | -3% |
| 60 | 730 | 81 | -17% |
| 70 | 820 | 86 | -17% |
| 80 | 893 | 119 | -17% |
| 90 | 945 | 151 | -17% |
| 100 | 966 | 174 | -21% |

- **CPPI performance**

To study the performance of CPPI strategy in our proposed model, we used daily prices of 50 securities in Tehran Stock Exchange from late March 2015 to early July 2015. Considering initial wealth of 100,000 USD, and the floor value of 70,000 USD, the uncertain portfolio rebalancing problem is solved continuously and the best portfolio is selected in each period. In addition, for better understanding of CPPI performance, we also solve the model by buy-and-hold (B&H) strategy – considering monthly rebalancing – and the results corresponding to the final wealth after rebalancing are compared. Moreover, to better realize the effects of each strategy on model, we simulate hypothetical bull market and bear market situations and the portfolio rebalancing model is also solved considering these situations.

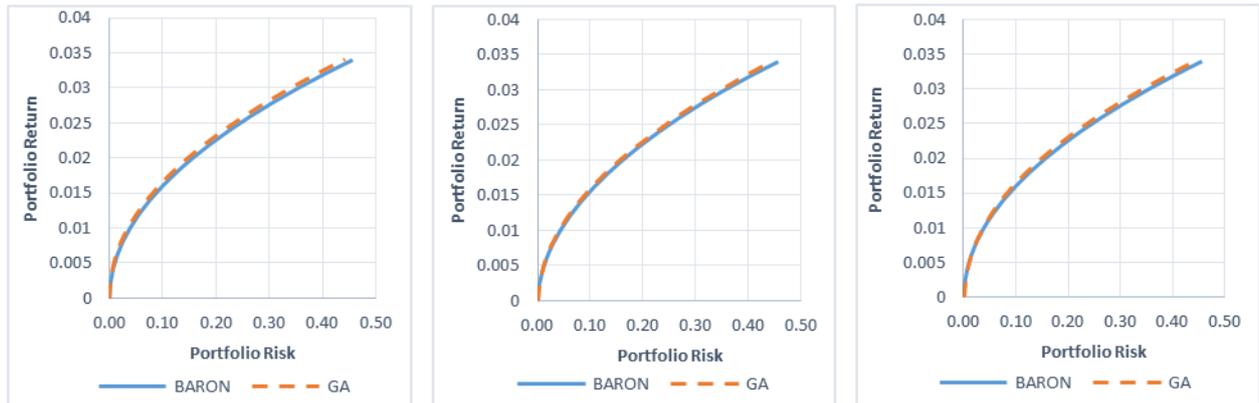

**Figure 2.** Efficient frontiers obtained by BARON solver vs. GA, for small problem with 10 risky assets (Left), with 15 risky assets (Middle), and with 20 risky assets (Right)

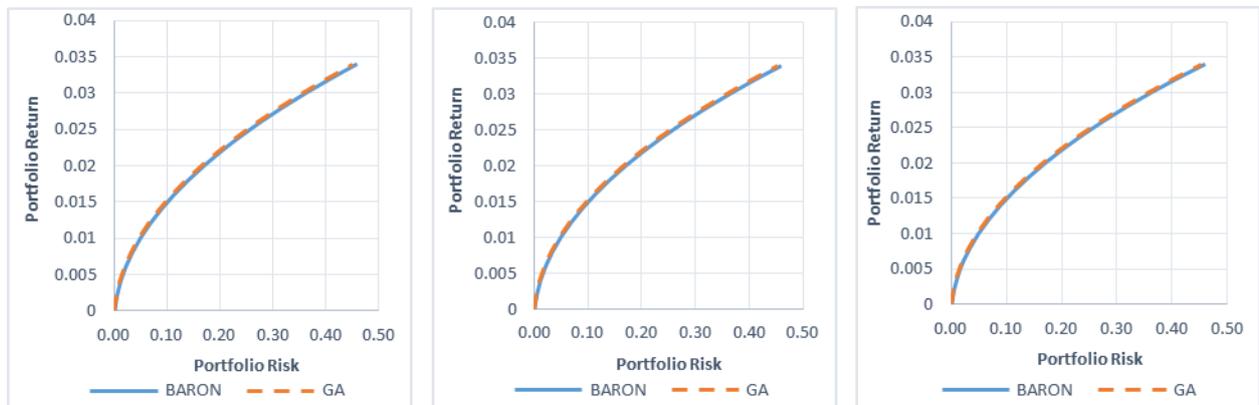

**Figure 3.** Efficient frontiers obtained by BARON solver vs. GA, for medium problem with 30 risky assets (Left), with 40 risky assets (Middle), and with 50 risky assets (Right)

Figure 5 (Left) illustrates CPPI performance versus B&H in flat market. Since CPPI allocates more money to risky assets as the total wealth increases, there will be an upside potential for CPPI strategy by increasing stock prices. In meantime, if there is a drop in stock prices (in which CPPI has allocated more money to), there will be more reduction on wealth level for CPPI than for B&H. This can be observed in Figure 5 (Left) as the wealth level on CPPI lowers into the B&H wealth level.

Figure 5 (Middle) illustrate similar situation for the total wealth level in a bear market. It can be seen that CPPI better controls risk than B&H as the prices fall. Although allocating more money to stocks at first has caused CPPI wealth levels to decline faster in compare with B&H, by gradually transferring money to risk-free bonds, CPPI insures that the wealth will never fall below the floor of 70,000.

Figure 5 (Right) compares the two strategies in a bull market. As the prices rise, it can be observed that CPPI has more potential in gaining profit and increasing the portfolio return, and consequently increasing the total wealth.

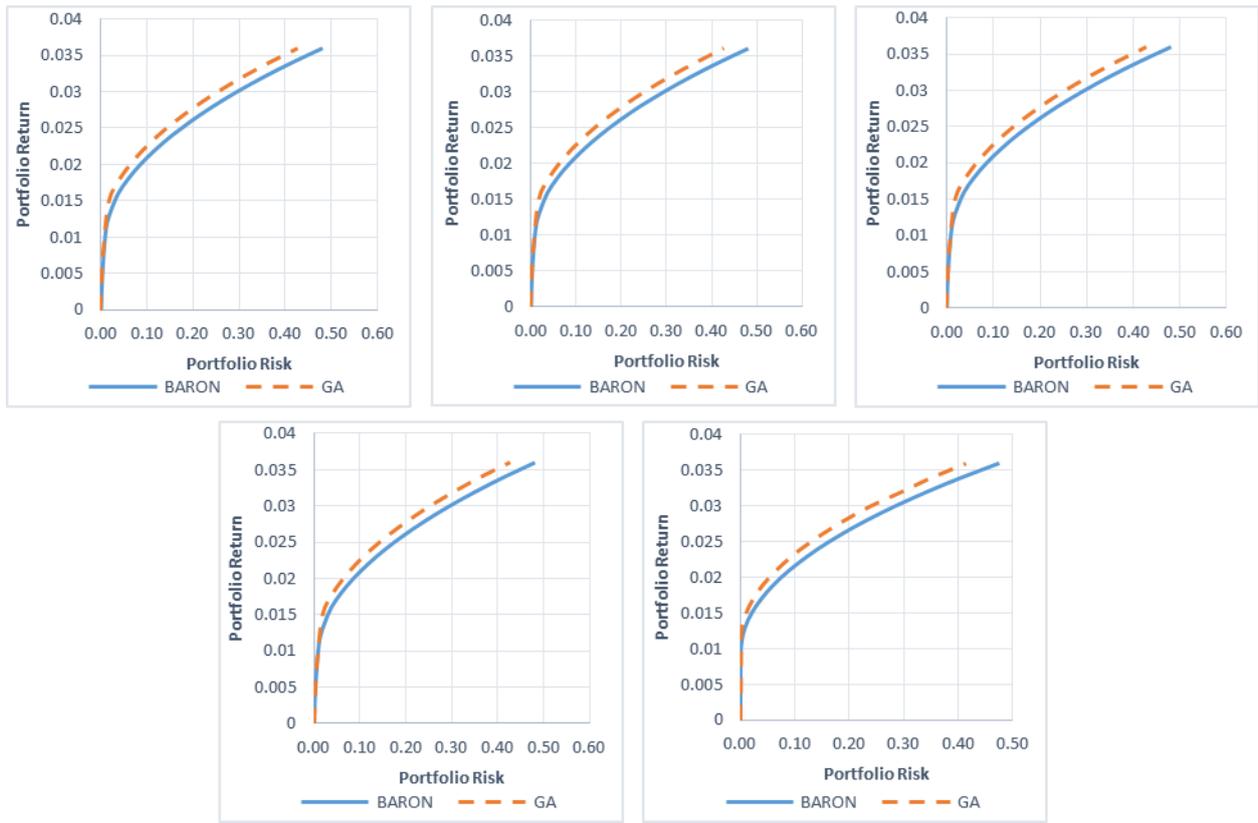

**Figure 4.** Efficient frontiers obtained by BARON solver vs. GA, for large problem with 60 risky assets (Up-Left), with 70 risky assets (Up-Middle), with 80 risky assets (Up-Right), with 90 risky assets (down-Left), and with 100 risky assets (Down-Right)

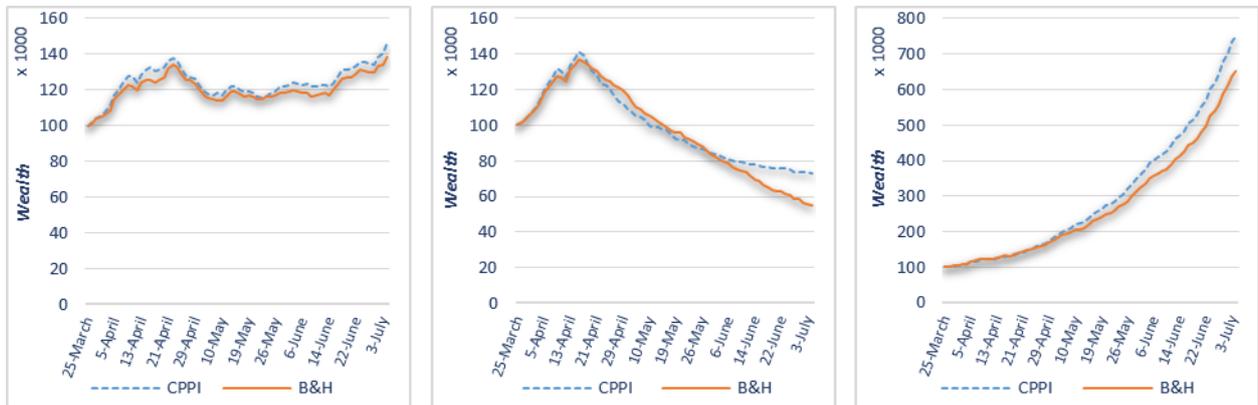

**Figure 5.** Performance of CPPI vs. B&H in changing portfolio value (total wealth) in Flat market (Left), Bear market (Middle), and Bull market (Right)

- **Belief degrees effect**

Belief degrees are results of experts' subjective evaluations, and therefore different people might produce different belief degrees. That is why it is important to investigate the impact of changes in belief degrees in proposed uncertain portfolio rebalancing problem. For this purpose, we consider six different levels for belief degrees on 10 risky assets where from level 1 to 6, the belief degrees become more conservative and contain wider ranges. Normal uncertainty distributions estimated for each asset in each level are shown in Table 7. The *zeroth* asset in each level represents the risk-free participation bonds. The uncertain portfolio rebalancing problem is solved for each level. Figure 6 illustrates the efficient frontiers corresponding to each level. According to the efficient frontiers it can be observed that in a constant risk level we will gain lower returns as the belief degrees become wider (from level 1 to level 6), which means that experts' evaluations or their preferences have been more conservative.

## 6. Conclusion and future research

In this paper, the portfolio rebalancing problem was modeled considering uncertain variables, transaction costs and constant-proportion portfolio insurance (CPPI) as rebalancing strategy. Our proposed model was solved using BARON Solver in GAMS and Genetic Algorithm in MATLAB. BARON is a computational system for solving non-convex optimization problems to global optimality. Solving numerical examples with real data showed that BARON does not guarantee global optimality, not even after finding 20 feasible solutions for each example, and the solver is terminated by reaching its time limit. Therefore, it is recommended that a meta-heuristic algorithm such as GA be utilized in order to reduce processing time and to obtain better solutions especially in large size problems. The results showed that the proposed algorithm in this paper performs well. In compare with BARON solver, it reduced the processing time by 90% on average, and improved the portfolio risk by 2%, 6% and 18% for small, medium and large problems, respectively. Moreover, the efficient frontiers (Pareto solutions) obtained by GA are more preferable, especially for problems with large sizes.

On the other hand, considering different levels of belief degrees confirmed changing in model solutions and efficient frontier diagrams. Therefore, considering uncertain variables affects portfolio rebalancing model and wider belief degree ranges make more conservative results. Furthermore, CPPI strategy performs as an insurance mechanism and limits downside risk in bear markets while it allows potential benefit in bull markets. Therefore, CPPI strategy has better performance than buy-and-hold strategy in portfolio rebalancing problem especially in bearish and bullish markets.

Future researches can include liquidity, price volatility, jumps in asset prices, the possibility of loans and short selling in their model in order to better illustrate real market situations. In addition, in order to improve the performance of the algorithm, future researches can consider other meta-heuristics such as artificial bee colony algorithm. Furthermore, it is recommended to study the performance of the model in other markets such real states. Finally, considering dynamic values for multiplier $m$ in CPPI strategy, discrete-time CPPI, and applying other strategies including OBPI in uncertain portfolio rebalancing problem is suggested for future studies.

**Table 7.** Estimated uncertainty distributions of returns on each asset for different belief degree levels

| Asset $i$ | Level 1 | Level 2 | Level 3 |
|---|---|---|---|
| 0 | Constant ≡ 0.00056 | Constant ≡ 0.00056 | Constant ≡ 0.00056 |
| 1 | $\mathcal{N}(0.00045, 0.02776)$ | $\mathcal{N}(0.00045, 0.03053)$ | $\mathcal{N}(0.00045, 0.03331)$ |
| 2 | $\mathcal{N}(0.00104, 0.01516)$ | $\mathcal{N}(0.00104, 0.01668)$ | $\mathcal{N}(0.00104, 0.01819)$ |
| 3 | $\mathcal{N}(0.00078, 0.01914)$ | $\mathcal{N}(0.00078, 0.02105)$ | $\mathcal{N}(0.00078, 0.02297)$ |
| 4 | $\mathcal{N}(0.00075, 0.02502)$ | $\mathcal{N}(0.00075, 0.02752)$ | $\mathcal{N}(0.00075, 0.03003)$ |
| 5 | $\mathcal{N}(0.00045, 0.01608)$ | $\mathcal{N}(0.00045, 0.01769)$ | $\mathcal{N}(0.00045, 0.0193)$ |
| 6 | $\mathcal{N}(0.06113, 1.01373)$ | $\mathcal{N}(0.06113, 1.11511)$ | $\mathcal{N}(0.06113, 1.21648)$ |
| 7 | $\mathcal{N}(0.00148, 0.02443)$ | $\mathcal{N}(0.00148, 0.02688)$ | $\mathcal{N}(0.00148, 0.02932)$ |
| 8 | $\mathcal{N}(0.00021, 0.0188)$ | $\mathcal{N}(0.00021, 0.02068)$ | $\mathcal{N}(0.00021, 0.02256)$ |
| 9 | $\mathcal{N}(-0.00025, 0.01858)$ | $\mathcal{N}(-0.00025, 0.02044)$ | $\mathcal{N}(-0.00025, 0.0223)$ |
| 10 | $\mathcal{N}(-0.00173, 0.01285)$ | $\mathcal{N}(-0.00173, 0.01413)$ | $\mathcal{N}(-0.00173, 0.01542)$ |
| Asset $i$ | Level 4 | Level 5 | Level 6 |
| 0 | Constant ≡ 0.00056 | Constant ≡ 0.00056 | Constant ≡ 0.00056 |
| 1 | $\mathcal{N}(0.00045, 0.03803)$ | $\mathcal{N}(0.00045, 0.03969)$ | $\mathcal{N}(0.00045, 0.04497)$ |
| 2 | $\mathcal{N}(0.00104, 0.02077)$ | $\mathcal{N}(0.00104, 0.02168)$ | $\mathcal{N}(0.00104, 0.02456)$ |
| 3 | $\mathcal{N}(0.00078, 0.02622)$ | $\mathcal{N}(0.00078, 0.02737)$ | $\mathcal{N}(0.00078, 0.031)$ |
| 4 | $\mathcal{N}(0.00075, 0.03428)$ | $\mathcal{N}(0.00075, 0.03578)$ | $\mathcal{N}(0.00075, 0.04054)$ |
| 5 | $\mathcal{N}(0.00045, 0.02203)$ | $\mathcal{N}(0.00045, 0.023)$ | $\mathcal{N}(0.00045, 0.02605)$ |
| 6 | $\mathcal{N}(0.06113, 1.38882)$ | $\mathcal{N}(0.06113, 1.44964)$ | $\mathcal{N}(0.06113, 1.64225)$ |
| 7 | $\mathcal{N}(0.00148, 0.03347)$ | $\mathcal{N}(0.00148, 0.03494)$ | $\mathcal{N}(0.00148, 0.03958)$ |
| 8 | $\mathcal{N}(0.00021, 0.02576)$ | $\mathcal{N}(0.00021, 0.02689)$ | $\mathcal{N}(0.00021, 0.03046)$ |
| 9 | $\mathcal{N}(-0.00025, 0.02546)$ | $\mathcal{N}(-0.00025, 0.02657)$ | $\mathcal{N}(-0.00025, 0.0301)$ |
| 10 | $\mathcal{N}(-0.00173, 0.0176)$ | $\mathcal{N}(-0.00173, 0.01837)$ | $\mathcal{N}(-0.00173, 0.02081)$ |

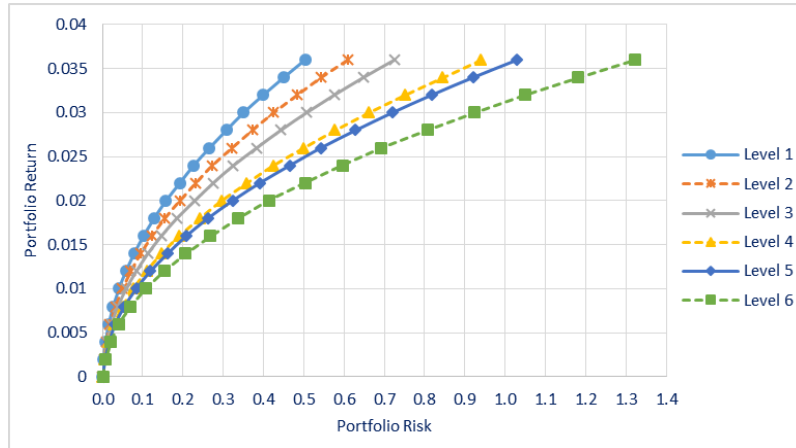

**Figure 6.** Efficient frontiers corresponding to different levels of belief degrees. The more conservative the belief degrees are evaluated, the lower the efficient frontier curve will be.

## Biographies

**Mostafa Zandieh** accomplished his BSc in Industrial Engineering at Amirkabir University of Technology, Tehran, Iran (1994-1998), and MSc in Industrial Engineering at Sharif University of Technology, Tehran, Iran (1998-2000). He obtained his PhD in Industrial Engineering from Amirkabir University of Technology, Tehran, Iran (2000-2006). Currently, he is an Associate Professor at Industrial Management Department, Shahid Beheshti University, Tehran, Iran. His research interests are production planning and scheduling, financial engineering, quality engineering, applied operations research, simulation, and artificial intelligence techniques in the areas of manufacturing systems design.

**Seyed Omid Mohaddesi** received his BSc in Industrial Engineering at University of Tabriz, Tabriz, Iran (2008-2012), and MSc in Financial Engineering at Raja University, Qazvin, Iran (2013-2015). Currently, he opts to pursue his PhD at a highly ranked university. His research interests include operations research, risk management, asset pricing, decision making, scheduling, and artificial intelligence techniques.